\documentclass[aps,prb,twocolumn,amssymb,showpacs,superscriptaddress]{revtex4}
\usepackage{graphicx}
\usepackage{amsmath}
\usepackage{color}
\newcommand{\tinyonehalf}{\frac{\mbox{\tiny 1}}{\mbox{\tiny 2}}}
\def\tone{\hat{1}}
\def\t1{{\hat{\tau}_1}}
\def\t2{{\hat{\tau}_2}}
\def\t3{{\hat{\tau}_3}}

\newcommand{\hv}{\hat{v}}
\newcommand{\hg}{\hat{g}}
\newcommand{\hG}{\hat{G}}
\newcommand{\chG}{\check{G}}
\newcommand{\chg}{\check{g}}

\newcommand{\grad}{\mbox{\boldmath$\nabla$}}

\def\vx{{\bf x}}
\def\vr{{\bf r}}

\def\vv{{\bf v}}
\def\vp{{\bf p}}
\def\vR{{\bf R}}
\def\vj{{\bf j}}

\def\vA{{\bf A}}
\newcommand{\hHam}{{\hat{\mbox{\footnotesize H}}}}

\newcommand{\Sig}{{\mbox{\footnotesize $\Sigma$}}}
\newcommand{\hSig}{{\hat{\mbox{\footnotesize $\Sigma$}}}}

\newcommand{\Area}{{\cal A}}
\begin{document}
\title{Shot noise in normal metal|$d$-wave superconducting junctions}
\author{Tomas L\"ofwander}
\email{tomas@snowmass.phys.nwu.edu} \affiliation{Department of Physics \& Astronomy, Northwestern University,
Evanston, IL 60208}
\author{Mikael Fogelstr\"om}
\affiliation{Department of Physics \& Astronomy, Northwestern University, Evanston, IL 60208}
\affiliation{Department of Theoretical Physics,
             Chalmers University of Technology and G\"oteborg University,
             S-412 96 G\"oteborg, Sweden}
\author{J. A. Sauls}
\affiliation{Department of Physics \& Astronomy, Northwestern University, Evanston, IL 60208}
\date{\today}
\begin{abstract}
We present theoretical calculations and
predictions for the shot noise in voltage
biased junctions of $d_{x^2-y^2}$
superconductors and normal metal
counter-electrodes. In the clean limit for the
d-wave superconductor the shot noise vanishes
at zero voltage because of resonant Andreev
reflection by zero-energy surface bound states.
We examine the sensitivity of this resonance to
impurity scattering. We report theoretical
results for the magnetic field dependence of
the shot noise, as well the fingerprints of
subdominant $s$- and $d_{xy}$ pairing channels.
\end{abstract}
 \maketitle
\section{Introduction}

It is now widely accepted that the order parameter in the superconducting state
of the cuprates has $d_{x^2-y^2}$ symmetry. Several experiments, including the
tri-crystal ring experiments,\cite{tsu00} corner junction
experiments,\cite{van95} and the $c$-axis STM probes of impurity
states,\cite{hud99,yaz99,pan00} have all provided strong evidence for
the broken reflection symmetry of the $d_{x^2-y^2}$ order parameter.

Experiments based on tunnelling within the $ab$-plane are another class of
experiments which provide valuable information about the properties of the
cuprates. Conductance measurements on planar or point contact (by STM)
normal metal|high-$T_c$ superconducting junctions, and on Josephson
junctions (e.g. grain boundaries) probe the electronic states of the
cuprates near the surface or at the interface. In addition to changes in
the atomic-scale structure of the interface, the $d_{x^2-y^2}$
superconducting state is highly distorted by interface scattering and
disorder. The electronic spectrum is strongly modified, and the
$d_{x^2-y^2}$ order parameter is in general suppressed on the coherence
length scale. In fact, a standard feature of the $ab$-plane
conductance is a large zero-bias conductance peak
(ZBCP).\cite{gee88,les92,cov97} Its sensitivity to impurity scattering and
the splitting of the peak in a magnetic field agree well with theoretical
predictions\cite{fog97a,poe00,fog03c} of surface Andreev bound states with
large spectral weight that provide a resonant channel for tunnelling near
zero bias. Normal scattering of quasiparticles by the surface from a
positive lobe of the $d_{x^2-y^2}$ order parameter to a negative lobe,
combined with Andreev reflection by the sudden $\pi$ phase shift (sign
change), leads to a zero-energy bound
state.\cite{hu94,buc95b,tan95} In the case of a specular [110] surface all
trajectories for quasiparticles are associated with a sign
change of the order parameter, and thus the spectral weight of
the Andreev states is very large.

In this paper we investigate theoretically the charge current through
voltage-biased normal metal--insulating-barrier--$d$-wave superconductor
(NIS) junctions subject to an external magnetic field directed along the
$c$-axis. We extend the theory for the current fluctuations of
conventional NIS junctions by Khlus\cite{khl87} to voltage-biased NIS
junctions with unconventional pairing correlations, as well as the effects
of field-induced surface currents on the current fluctuations. We present
calculations of both the mean charge current and the charge current
fluctuations, which at zero temperature reduce to the shot noise. We show
that the resonant nature of Andreev reflection via the surface bound
states in a clean $d_{x^2-y^2}$ superconductor can be used to extract additional
information from the shot noise that cannot be obtained from conductance
measurements alone. We discuss how the surface Andreev bound states
(ABS) are broadened by impurity scattering. The impurity effect is
reflected both in the conductance, which is related to the local density
of states at the interface, and the shot noise, which reduces to $S=2eI$
when impurity scattering dominates the intrinsic width of the surface ABS.

There are theoretical reasons to expect an additional, subdominant
pairing state, e.g. with $s$- or $d_{xy}$ symmetry, to be present in
equilibrium under favorable circumstances.\cite{sig95,buc95a,mat96,fog97a,gra00}
The formation of surface states at the Fermi level, in combination with an attractive,
sub-dominant pairing interaction favors a mixed-symmetry order
parameter, e.g. a surface phase with $d_{x^2-y^2}\pm is$ or $d_{x^2-y^2}\pm
id_{xy}$ symmetry. The sub-dominant component is predicted to have a
phase of $\pm\pi/2$ relative to the dominant $d_{x^2-y^2}$ component, and thus
to exhibit spontaneously broken time-reversal symmetry ($\textsf{T}$-symmetry).
The internal phase of the order parameter leads to a shift of the bound state
energies below the Fermi level, thus lowering the surface free energy and generating
a spontaneous surface current.\cite{mat96,fog97a}

The splitting of the Andreev states also leads to the prediction that
the ZBCP should spontaneously split as a function of
voltage.\cite{fog97a} Such a splitting has been observed in
\texttt{YBCO} near optimal doping.\cite{cov97,kru99} However, in
contrast to the general consensus about the $d_{x^2-y^2}$ pairing
symmetry for the bulk phase of the cuprates, the nature of the surface
phase, including the possibility of broken $\textsf{T}$-symmetry, is
unsettled.\cite{nei02,fog03c} In the following we also discuss the
``fingerprints'' of sub-dominant pairing that may be observable in the shot
noise.

The theory of shot noise in mesoscopic, metallic systems has been used in several
recent experiments to gain information about tunnelling in mesoscopic systems, see e.g.
the recent review in Ref.~\onlinecite{but00}.
Earlier predictions for shot noise in $d$-wave NIS junctions were published by Zhu and
Ting.\cite{zhu99} It was shown in a non-self-consistent calculation that for a clean
d-wave superconductor the Andreev resonance associated with the surface Andreev bound
states suppressed the shot noise to zero. It was confirmed in
Ref.~\onlinecite{tan00} that this effect also holds when self-consistency of
the order parameter is taken into account. Recent theoretical work on
noise in d-wave SIS junctions addresses the effects of surface disorder,
interface states and magnetic fields on multiple Andreev reflection in this
system, which is an important mechanism for noise in voltage-biased Josephson
junctions.\cite{cue01,cue02}

Here we report a detailed study of the shot noise, and specifically address physical
conditions that are likely present at an NIS interface. The shot noise is shown to be
particularly sensitive to the spectrum of surface states and to disorder. We present
results for the magnetic field dependence of the shot noise, and demonstrate the
sensitivity of the results obtained for clean d-wave superconductors to changes in the
low-energy electronic spectrum by equilibrium surface currents and impurity scattering.
We focus on the [110] orientation of an NIS interface to the cuprate superconductor,
since for this orientation the influence of the Andreev bound states is
most pronounced.

In Section~\ref{SecModel} we describe our model of the normal metal-unconventional
superconductor contact, beginning with a brief review of the quasiclassical Green's
function technique that we use to compute observables. We discuss the boundary
conditions for the non-equilibrium propagators, coherence amplitudes and distribution
functions, and express these boundary conditions
in terms of generalized scattering amplitudes. Explicit solutions are used to
obtain results for the transport current and spectral density for current noise
under nonequilibrium steady-state conditions.
In Section~\ref{SecNoise} we present the results for the shot noise in
voltage-biased NIS junctions with d-wave pairing symmetry for junctions
with disorder near the interface. We discuss the effects of magnetic fields
and screening currents and surface phase transitions on the noise spectrum.
Throughout the paper we use units for which
$\hbar=k_B=1$, and we choose the sign convention $e=-|e|$.

\section{Theory and Interface Model}\label{SecModel}

To compute the average current and fluctuations of the current we use the
quasiclassical Green's function method,\cite{ser83,lan86} and the Keldysh
formalism to calculate non-equilibrium properties. The relevant
information about the spectrum of current-carrying states and their
distribution functions are contained in a set of non-equilibrium matrix
Green's function: the retarded (R), advanced (A) and Keldysh (K)
propagators, $\hg^{R,A,K}(\vp_f,\vR;\epsilon,t)$, which are $4\times 4$
matrix propagators in the combined spin and particle-hole space (Nambu
space), that depend on the Fermi momentum, $\vp_f$, the excitation
energy, $\epsilon$, and space and time coordinates, $\vR$ and $t$. The
quasiclassical propagators are related to the full Nambu propagators:
\begin{eqnarray}
\hG^R(x,x')&=&-i\Theta(t-t')\langle\{\Psi(x)\,,\,\bar\Psi(x')\}\rangle\,,\nonumber\\
\hG^A(x,x')&=&+i\Theta(t'-t)\langle\{\Psi(x)\,,\,\bar\Psi(x')\}\rangle\,,\nonumber\\
\hG^K(x,x')&=&-i\langle\left[\Psi(x)\,,\,\bar\Psi(x')\right]\rangle
\,,
\end{eqnarray}
where the Nambu field operators,
$\Psi(x)=(\psi_{\uparrow}(x),\psi_{\downarrow}(x),
\psi^{\dagger}_{\uparrow}(x),\psi^{\dagger}_{\downarrow}(x))^{\text{\tiny tr}}$,
and $\bar\Psi(x)=\Psi^{\dagger}(x)$ incorporate particle-hole coherence
of the superconducting state. We use the notation defined in Ref. \onlinecite{ser83}
for the two-point functions where $\Theta(t)$ is the Heavyside function,
$\{A,B\} = AB+BA$, and $[A,B]=AB-BA$. We also use the short-hand notation $x=(\vx,t)$.

A compact formulation of the non-equilibrium equations is
obtained in the Keldysh formulation in which the set
Nambu-matrix propagators, $\hG^{R,A,K}$, are grouped
into a $2\times 2$ Keldysh matrix,
\begin{equation}
\hspace*{-2mm}\chG(\vp,\vR;\epsilon,t)=\int dr\,e^{-i(\vp\cdot\vr-\epsilon\tau)}\,
\left(\begin{array}{cc}\hG^R & \hG^K \\ 0 & \hG^A\end{array}\right)
\,.
\end{equation}
It is most convenient to work in terms of the center-of-mass and relative
coordinates, $R=(x+x')/2=(\vR,t)$ and $r=x-x'=(\vr,\tau)$, and Fourier transform
with the relative space and time coordinates.
The quasiclassical propagators are then defined in terms of an integration of the full
Keldysh-Nambu matrix propagator, $\chG(\vp,\vR;\epsilon,t)$, over an energy shell
that is small compared with the Fermi energy,
$|v_f(p-p_f)|<\varepsilon_c\ll E_f$,
\begin{equation}
\chg(\vp_f,\epsilon;\vR,t)=\frac{1}{a}
\int_{-\varepsilon_c}^{+\varepsilon_c}
d\xi_{\vp}\,\check{\tau}_3\chG(\vp,\epsilon;\vR,t)
\,.
\end{equation}
The quasiclassical propagator is defined by dividing out the weight of the quasiparticle
pole in the spectral function, $a$, and by convention pre-multiplying by the matrix,
$\check\tau_3=\hat\tau_3\check 1$. We denote a Nambu matrix with a `hat',
while Keldysh matrices are denoted with a `check'. Thus, $\hat\tau_3$ is the third Pauli
matrix in the particle-hole sector of Nambu space, and $\check{1}$ is the identity
Keldysh matrix.

For pure spin-singlet pairing considered here the quasiclassical propagators,
$\hg^{R,A,K}$, may be parameterized in particle-hole space by scalar amplitudes
for the diagonal (quasiparticle) and off-diagonal (Cooper pair) propagators,
\begin{equation}
\hg^{X} =
\left(\begin{array}{cc} g^X & f^X \\ \underline{f}^X & \underline{g}^X\end{array}\right)
\,,
\label{eq:gmt}
\end{equation}
where $X=(R,A,K)$. We consider the case where the diamagnetic coupling of the charge
currents to the magnetic field dominates the paramagnetic Zeeman coupling, in which
case the spin degrees of freedom are inert. The components of the quasiclassical propagators
are then defined in terms of spin scalar diagonal propagators, $g^X$ and $\underline{g}^X$,
and spin-singlet off-diagonal propagators, $f^X$ and $\underline{f}^X$.
These components are not all independent, but are related by symmetries that follow from
the fermion anti-commutation relations.\cite{ser83}

For calculating the low-frequency ($\omega\ll \Delta,eV$) conductance and
noise in NIS tunnel junctions we need only time-independent propagators. The
steady-state nonequilibrium quasiclassical Keldysh-matrix propagator obeys
a matrix transport equation,
\begin{equation}\label{transport_eq}
\left[\epsilon\check\tau_3-\check{v}-\check\Sig\,,\,\chg \right]
+ i\vv_f\cdot\grad\chg = 0
\,,
\end{equation}
where $\check v = e\Phi \check{1}$ is the electrostatic potential and
$\check\Sig=\check\Delta+\check\Sig_{\text{\tiny imp}}$ represents the order
parameter and impurity self energy. The transport equation is supplemented by
the normalization condition, $\chg^2 = -\pi^2\check{1}$,\cite{eil68,lar68} and
by boundary conditions connecting the propagators at the interface. When there
is no reason for confusion, we do not display the dependence of $\check g$ and
$\check\Sig$ on $(\vp_f,\vR;\epsilon)$. Equation (\ref{transport_eq}) represents
coupled equations for the retarded, advanced,
\begin{equation}
\label{QC_RA}
\left[\hHam^{R,A}\,,\,\hg^{R,A}\right]+i\vv_f\cdot\grad\hg^{R,A} = 0
\,,
\end{equation}
and Keldysh propagators,
\begin{eqnarray}
\label{QC_K}
\hHam^{R}\hg^{K}-\hg^{K}\hHam^{A}&+&\hg^{R}\hSig^{K}-\hSig^{K}\hg^{A}
\nonumber\\
&+&i\vv_f\cdot\grad\hg^{K} = 0
\,,
\end{eqnarray}
where $\hHam^{R,A}=\epsilon\t3-\hv-\hSig^{R,A}$
is defined in terms of the excitation energy, $\epsilon$, the coupling
to external fields, $\hv$, and the self-energies, $\hSig^{R,A}$.
Similarly, the normalization condition expands to
\begin{equation}\label{normRAK}
\hg^{R,A}\,\hg^{R,A} = -\pi^2\,\tone
\,,\quad
\hg^{R}\,\hg^{K} - \hg^{K}\,\hg^{A} =0
\,.
\end{equation}
The retarded and advanced functions determine the particle-hole coherence functions
and spectral properties at the NIS interface, while the Keldysh function contains additional
information on the non-equilibrium distribution of these states.

The computation of $\chg$ involves solving the quasiclassical transport equations for
the normal metal-insulating barrier-superconductor system together with a set of
self-consistency equations for the impurity and pairing self-energies
and boundary conditions in the bulk reservoirs and
at the interface.

\subsection{Pairing Model}

The pairing correlations are described by the pairing self-energy, $\check\Delta$.
In the leading order (weak-coupling) approximation, the Keldysh component vanishes
and the retarded and advanced self-energies are independent of energy. The resulting
self-consistency condition is the BCS gap equation,
\begin{equation}\label{BCS_gap-equation}
\hspace*{-2mm}\hat\Delta(\vp_f,\vR)=
\Big\langle\lambda(\vp_f,\vp_f')\,
\int_{-\epsilon_c}^{+\epsilon_c}\frac{d\epsilon}{4\pi i}\hat f^K(\vp_f',\vR;\epsilon)
\Big\rangle_{\vp_f'}
\,.
\end{equation}
where $\hat f^K (\vp_f,\vR;\epsilon)$ is the off-diagonal
quasiclassical Keldysh propagator.
We consider spin-singlet superconductivity derived from a pairing interaction of
the form,
\begin{equation} \label{interaction}
\lambda(\vp_f,\vp_f')=\sum_{\alpha}\lambda_{\alpha}
                      \eta_{\alpha}(\vp_f)
                      \eta_{\alpha}(\vp_f')
\,,
\end{equation}
where the sum is over the `relevant' irreducible representations of the
crystal point group, $D_{4h}$;
$\alpha\in\{
A_{\text{\tiny 1g}}(s-\text{wave})
B_{\text{\tiny 1g}}(d_{x^2-y^2}-\text{wave}),
B_{\text{\tiny 2g}}(d_{xy}-\text{wave}),
A_{\text{\tiny 2g}}(g-\text{wave})
\}$,
and $\lambda_{\alpha}$ and $\eta_{\alpha}(\vp_f)$ are the corresponding
eigenvalues and eigenfunctions for pairing in channel $\alpha$. The `relevant' channels
are defined by the dominant attractive eigenvalues for each
irreducible representation obtained from solutions of the linearized
gap equation with the microscopic pairing interaction
(c.f. Ref. \onlinecite{buc95a}). The mechanism for pairing in the
cuprates is not a solved problem, and indeed there may be more than
one mechanism at work in the cuprates account for the wide range of
superconducting properties as a function of doping and disorder.
For example, a relatively simple two-channel model based on electronic
coupling to anti-ferromagnetically correlated spin-excitations and to phonons leads to
dominant $d_{x^2-y^2}$ pairing over a wide range of doping, but with
significant sub-dominant pairing interactions in all other symmetry
channels.\cite{fog03c} These sub-dominant pairing channels are predicted
to play an important role in the local electronic structure of surface
superconducting state near an insulating barrier or other
interface.\cite{mat95,buc95a}

In this paper we consider the signatures of sub-dominant pairing in the shot
noise. For this purpose we assume the dominant pairing channel has
$d_{x^2-y^2}$ symmetry, and consider sub-dominant pairing in the $s$- or
$d_{xy}$ pairing channels, i.e. $\lambda_{B_{1g}} >
\lambda_{B_{2g}}\,\,\lambda_{A_{1g}}$. The $A_{2g}$ channel may also have an
attractive eigenvalue for spin-fluctuation dominant pairing, but this
order parameter is suppressed on both $[110]$ and $[100]$ boundaries, and is
particularly sensitive to surface disorder, so we do not consider this
sub-dominant channel for NIS junctions.

Below the superconducting transition temperature, $T_c$, the order parameter
amplitude is proportional to the $B_{1g}$ basis function,
$\eta_{B_{1g}}=\sqrt{2}(\hat p_x^2-\hat p_y^2)$. However, even a small
attractive sub-dominant eigenvalue, $\lambda_{B_{2g}}$ or $\lambda_{A_{1g}}$
can, at low temperature generate a transition to a state with a mixed
symmetry, with an order parameter that acquires an additional component
proportional to the corresponding eigenfunction, $\eta_{B_{2g}}=\sqrt{2}\hat
p_x\hat p_y$ or $\eta_{A_{1g}}=1$.\footnote{We use the simplest polynomials to
represent the basis functions. The true basis functions are obtained as
eigenfunctions of the linearized gap equation with true pairing interaction,
and in general differ from the polynomial functions we use here.}
Thus, in general we write the order parameter as
\begin{equation}\label{gap_ansatz}
\Delta(\vp_f,\vR)=\sum_{\alpha}\Delta_{\alpha}(\vR)\eta_{\alpha}(\vp_f)
\,.
\end{equation}
The gap equation separates into three self-consistency equations for
each relevant pairing channel,
\begin{equation}\label{gap_eqs}
\Delta_{\alpha}(\vR)=\lambda_{\alpha}
\Big\langle\eta_{\alpha}(\vp_f)
\int_{-\epsilon_{c}}^{\epsilon_{c}}\frac{d\epsilon}{4\pi i}f^K(\vp_f,\vR;\epsilon)
\Big\rangle_{\vp_f}
\,,
\end{equation}
The solution of the transport equation and boundary conditions lead to
coupling between the components, $\Delta_{\alpha}(\vR)$. The cutoff
$\epsilon_{c}$ and pairing interaction, $\lambda_{\alpha}$, are
eliminated in favor of the instability temperatures, $T_{c\alpha}$,
using the solution of the linearized gap equation,
$\lambda_{\alpha}^{-1}=\ln(T/T_{c\alpha})+ \int(d\epsilon/2\epsilon)
\tanh(\epsilon/2T)$. The overall phase of the order parameter
(\ref{gap_ansatz}) can be eliminated for an NIS system, but the
relative phases between the different components that remain are determined
by the minimum of the free energy. At sufficiently low temperature,
the lowest energy state near a [110] surface is always a mixed
symmetry phase with spontaneously broken $\textsf{T}$ symmetry, in
which the sub-dominant order parameter acquires a finite value with a
relative phase of $\pm\pi/2$.\cite{mat96,fog97a} Consequently, we
consider three possible order parameters: 1) pure $d_{x^2-y^2}$, 2)
$d_{x^2-y^2}+is$, and 3) $d_{x^2-y^2}+id_{xy}$. Cases 2 and 3 are illustrated
in Fig. ({\ref{Surface_OP}), where the pairing interaction
parameters are chosen so that the order parameter in the bulk region
is always pure $d_{x^2-y^2}$, and the subdominant components are stable
near the surface within a layer of the order of a few coherence lengths.
\begin{figure}[t]
\includegraphics[width=8cm]{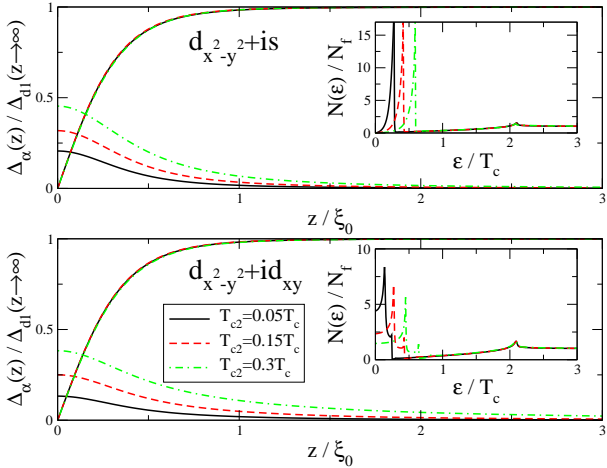}
\caption{\small Mixed-symmetry order parameters at a [110] specular surface.
The relative phase between the $d_{x^2-y^2}$ and sub-dominant $d_{xy}$ or
$s$ components is $\pm\pi/2$. The insets show the surface bound state spectrum
for these mixed-symmetry phases. The calculations were carried out in the
clean limit with $\Gamma=0.001T_c$ and $\sigma=10^{-4}$ (see text) at a
temperature $T=0.05T_c$.
}
\label{Surface_OP}
\end{figure}

\subsection{Magnetic Field and Screening Currents}

One of the key features of the ABS interpretation of the zero-bias conductance
peaks observed in ab-plane tunnelling spectroscopy is the splitting of the
ZBCP for low magnetic fields.\cite{les92,apr99,kru99} The energy of the ABS is
shifted away from the Fermi level by screening currents. The origin
of this effect is the coupling of the quasiparticle current to the
superflow field generated in response to the magnetic field,
\begin{equation}\label{Doppler}
\check v_{\text{\tiny A}} = \vv_f\cdot\vp_s\,\check\tau_3
\,,
\end{equation}
where the condensate flow field is given by the gauge-invariant gradient
of the phase, $\vp_s=\tinyonehalf(\grad\vartheta-\frac{e}{c}\vA)$, where
$\vA$ is the vector potential. We include this coupling here in order to
investigate the sensitivity of the noise spectrum to the spectral shift of
the surface ABS. Indeed as we show below the shot noise is particularly
sensitive to the Doppler shift of the zero energy surface states.

The condensate flow field, $\vp_s$, is calculated by solving
Maxwell's equation, self-consistently with the
surface current, supplemented with the boundary conditions for
magnetic field, $B\rightarrow 0$ for the Meissner state far
from the surface, and $B\rightarrow H_{\text{\tiny ext}}$ at
the surface. For strong type II superconductors, such as the cuprates,
with a magnetic penetration length $\lambda\gg\xi_0$,
the solution of Maxwell's equation to leading
order in the small parameter $\xi_0/\lambda$ is
$p_s(z)=p_{s0}e^{-z/\lambda}$ with
\begin{equation}\label{ps}
\frac{p_{s0}v_f}{T_c}=
\frac{H_{\text{\tiny ext}}}{H_0}-\frac{1}{\lambda}\int_0^{\infty}dz\,j_p(z)
\,,
\end{equation}
where $j_p(z)$ is the paramagnetic part of the current flowing parallel
to the interface [computed via Eq.~(\ref{current_full}) below]. The paramagnetic
current originates from the Doppler splitting of the ABS, which preferentially
favors the paramagnetic response from the bound states.\cite{fog97a}
The field scale in Eq. (\ref{ps}) is $H_0=\Phi_0/(\pi\xi_0\lambda)$,
where $\Phi_0=c/2|e|$ is the flux quantum.

\subsection{Impurity self energy}

The anisotropic order parameter (Eq. \ref{gap_ansatz}) is sensitive to
disorder. We include the leading order (in $1/p_f\ell_{\text{\tiny imp}}$
where $\ell_{\text{\tiny imp}}$ is the mean free path) effects of disorder
within the model of isotropic scattering of quasiparticles by impurities
(c.f. Ref. \onlinecite{gra96a}). In this model the impurity self-energy is
given by the quasiparticle-impurity $t$-matrix and the average impurity
concentration, $n_{\text{\tiny imp}}$,
\begin{equation}\label{Sigma_impurity}
\check\Sig_{\text{\tiny imp}}(\vp_f,\vR;\epsilon)=
  n_{\text{\tiny imp}}\check t({\vp_f,\vp_f,\bf R};\epsilon)
\,,
\end{equation}
where $\check t(\vp_f,\vp_f',\vR;\epsilon)$ satisfies a
Bethe-Salpeter equation for repeated scattering of quasiparticles
by impurities.\cite{gra96a} For isotropic impurity scattering defined
by a scattering amplitude, $u_0$, the t-matrix equations for the retarded
and advanced self-energies have the solutions,
\begin{equation}\label{t-matrix_RA}
\hat t^{R,A}(\vR;\epsilon) =
\frac{u_0\left[\hat 1+u_0 N_f\langle\hg^{R,A}(\vp_f,\vR;\epsilon)\rangle_{\vp_f}\right]}
     {\hat 1-\left[u_0 N_f\langle\hg^{R,A}(\vp_f,\vR;\epsilon)\rangle_{\vp_f}\right]^2}
\,,
\end{equation}
and the Keldysh component is given by
\begin{equation}\label{t-matrix_K}
\hat t^K=N_f\hat t^R\langle\hg^K\rangle_{\vp_f}\hat t^A
\,.
\end{equation}

The scattering amplitude, $u_0$, defines the $s$-wave scattering phase shift,
$\delta_0=\arctan (\pi N_f u_0)$, while the impurity concentration and normal-state
density of states define an energy scale $\Gamma_0=n_{imp}/(\pi N_f)$. We
use the more common parametrization of the impurity scattering model in terms
of the dimensionless scattering cross section, $\sigma=\sin^2\delta_0$, and the pair
breaking parameter, $\Gamma=\Gamma_0\sin^2\delta_0$, or equivalently the
mean-free path, $\ell_{\text{\tiny imp}}=v_f/2\Gamma$.

The impurity self-energy renormalizes the excitation spectrum via,
$\epsilon\rightarrow\tilde\epsilon(\vR)=\epsilon-\Sig^R_3(\epsilon,\vR)$,
where $\Sig^R_3$ is the $\hat\tau_3$-component of $\hSig^R_{\text{\tiny imp}}$.
For a pure $d_{x^2-y^2}$ pairing state, the impurity renormalization of the order parameter
vanishes by symmetry; the effects of pair-breaking are included through
the renormalized excitation spectrum. As a result the solution of the gap equation
shows a reduction of the order parameter amplitude with increasing pair breaking
parameter, $\Gamma$. For mixed symmetry pairing, e.g. at the surface, and in the
presence of field-induced screening currents,
the pairing self-energy is, in general, non-vanishing and must be calculated self-consistently
with the renormalization of the excitation spectrum.

In general the self-energy also includes electron correlation effects generated by
electron-electron and electron-phonon interactions. In what follows we consider a
simplified model for the metallic electrodes in which the quasiparticles are
governed by an effective one-electron Hamiltonian with a parabolic band structure.
Thus, the only electronic correlations included here are those that contribute to
the effective mass, $m^*$, and give rise to superconductivity. In this case,
$\vv_f=\vp_f/m^*$, and the charge current carried by a normal quasiparticle is
$e\vv_f=\frac{e}{m^*}\vp_f$. Both the current and the noise spectrum are then
calculated in this effective one-electron theory, modified to include pairing
correlations in the superconductor, effects of screening currents on
the surface excitation spectrum and impurity scattering in both electrodes.

\subsection{Current and Current-Current Correlations}

Physical properties, such as the local excitation spectrum, current response
and correlation functions are expressed in terms of the quasiclassical
Green's function. Here we are interested in computing the charge current
and the mean-field fluctuations of the current for the NS junction.
The junction current is an expectation value, in a nonequilibrium ensemble ($\rho$),
of the Heisenberg operator for the charge current,
\begin{equation}\label{CurrentOperator}\begin{split}
\textsf{J}({\bf r}_1,t_1) &= \lim_{x_2\rightarrow x_1} \frac{-e}{2m^*i}
\left(\nabla_1-\nabla_2\right)\\
&\quad \left[\psi_{\uparrow}^{\dagger}(x_2)\psi_{\uparrow}(x_1)
-\psi_{\downarrow}^{\dagger}(x_1)\psi_{\downarrow}(x_2)\right]
\,,
\end{split}\end{equation}
Fluctuations of the current are related to the current-current
correlation function, which is defined in terms of the
operator,
\begin{equation}\label{NoiseOperator}
\textsf{S}({\bf r},t,\tau)\equiv
\textsf{K}({\bf r},t,t+\tau) +
\textsf{K}({\bf r},t+\tau,t)
\,,
\end{equation}
\begin{widetext}
\begin{equation}\label{NoiseDefinition}
\begin{split}
\textsf{K}({\bf r}_1,t_1,{\tilde t}_1) &=
\left(\frac{e}{2m^*i}\right)^2 \lim_{\tilde {\bf r}_1\rightarrow{\bf
r}_1}
\lim_{\substack{x_2\rightarrow x_1\\
{\tilde x}_2\rightarrow {\tilde x}_1}} \big(\nabla_1-\nabla_2\big)
\big(\tilde\nabla_1-\tilde\nabla_2\big)\\
&\quad\left[\psi_{\uparrow}^{\dagger}(x_2)\psi_{\uparrow}(x_1)
-\psi_{\downarrow}^{\dagger}(x_1)\psi_{\downarrow}(x_2)\right]
\left[\psi_{\uparrow}^{\dagger}(\tilde x_2)\psi_{\uparrow}(\tilde x_1) -
\psi_{\downarrow}^{\dagger}(\tilde x_1)\psi_{\downarrow}(\tilde x_2)\right]
-\vj({\bf r}_1,t_1)\vj({\bf r}_1,\tilde t_1)
\,,
\end{split}
\end{equation}
\end{widetext}
where $\vj=\text{Tr}[\rho\,\textsf{J}]$ is the expectation value of current operator.
The observable noise is found by evaluating the average of
this operator over the statistical ensemble.

In what follows we consider the effects of scattering by point impurities (s-wave) in the
superconducting electrode with d-wave pairing. For the steady-state conductance and
current noise the statistical average for the noise reduces to the product of two-point
correlation functions. Vertex corrections to the current-current correlator vanish
in the above approximations, or originate from quantum interference effects or
coupling to collective modes and thus are higher order in $1/p_f\ell_{\text{\tiny imp}}$
or $1/p_f\xi_0$, and neglected here. Thus, after integration over the cross section
of the junction, $\Area$, we obtain the current noise, $S(z_1,t_1,{\tilde t}_1)$,
\begin{widetext}
\begin{equation}
\label{ExchangeNoise} \hspace*{-3mm}S =
\frac{1}{2}\left(\frac{e}{2m^*i}\right)^2 \lim_{ \substack{
{\bf r}_2\rightarrow{\bf r}_1 \\
\tilde{\bf r}_2\rightarrow\tilde{\bf r}_1\\
\tilde{\bf r}_1\rightarrow{\bf r}_1} } \int d^2r_{1\|}
\int d^2{\tilde r}_{1\|}
\left(\partial_{z_1}-\partial_{z_2}\right)
\left(\partial_{\tilde z_1}-\partial_{\tilde z_2}\right)
\mbox{Tr}\left\{
\hat G^<(\tilde{\bf r}_1,{\bf r}_2;\tilde t_1,t_1)
\hat G^>({\bf r}_1,\tilde{\bf
r}_2;t_1,\tilde t_1) \right\},
\end{equation}
\end{widetext}
where $\hat G^{\gtrless}=\hat G^K \pm (\hat G^R-\hat G^A)$ are the
particle ($<$) and hole ($>$) correlation functions.

We separate out the momentum component parallel to the junction using the inverse
Fourier transformation with
respect to the difference coordinate ${\bf r}_{\|}={\bf r}_{1\|}-{\bf r}_{2\|}$,
\begin{equation}\label{parallel_momentum}
\hspace*{-2mm}\chG({\bf r}_1,{\bf r}_2) =
\int \frac{d^2p_{\|}}{(2\pi)^2} e^{i{\bf p}_{\|}\cdot{\bf r}_{\|}}
\chG (z_1,z_2,{\bf p}_{\|},{\bf R}_{\|}) \,,
\end{equation}
and assume that the dependence on ${\bf R}_{\|}$ is slow on the scale of the coherence
length, i.e. locally planar; thus, we omit $\vR_{\|}$. Near the junction, incident and
scattered waves interfere on a scale given by the inverse Fermi momentum $p_f^{-1}$. We
then make the following \textsl{ansatz},\cite{zai84,khl87,mil88} which factors the
propagator into rapidly oscillating incident and reflected waves with wavenumbers, $\pm
p_{fz}$, and slowly varying two-point
envelope functions,
\begin{equation}
\label{Zaitsev_expansion} \check\tau_3 \chG(z_1,z_2) = \frac{1}{v_{fz}} \sum_{\nu\mu} \check
C_{\nu\mu}(z_1,z_2) e^{ip_{fz}(\nu z_1-\mu z_2)} \,,
\end{equation}
where $z$ is the coordinate normal to the interface. The sum is over {\it
direction} indices $\nu$ and $\mu$ which are $+1$ or $-1$, depending on the sign of
the projection of the Fermi momentum on the $z$-axis for incident or reflected
waves (see Fig.~\ref{junction}). When $\nu$ appears as an index of a function we
use a shorthand notation $\pm$ for $\pm 1$. The diagonal (in direction index space)
functions $\check C_{\nu\nu}$ are related to Shelankov's two-point quasiclassical
Green's functions.\cite{she85} In the limits, $z_2\rightarrow z_1{\pm}0$,
these components are related to the projection operators,
\begin{eqnarray}
\mp i\check C_{\substack{++\\--}}=\check P_{\mp}=
    \frac{1}{2}\left(\check{1}\mp\frac{\chg}{-i\pi}\right)
\,.
\end{eqnarray}
The functions $\check C_{\nu(-\nu)}$ are the pre-factors of the rapidly oscillating
carrier waves, $\sim e^{\pm 2ip_{fz} z}$. These amplitudes are \textsl{drones} - slaved
to the quasiclassical propagators and ultimately eliminated from the boundary conditions
and observables such as the current noise. In particular, Zaitsev's boundary
conditions\cite{zai84} are derived (for details see Ref.~\onlinecite{mil88}) by
eliminating the drone amplitudes from a set of linear relations connecting the
quasiclassical projectors, $\check C_{\nu\nu}$ and the quasiclassical drones, $\check
C_{\nu(-\nu)}$. This procedure transforms the linear boundary conditions expressed in
terms of the set, $\{\check C_{\nu\mu}\}$, into a set of non-linear boundary conditions
for the quasiclassical projectors, $\check C_{\nu\nu}$, and consequently for the
quasiclassical Green's functions.

For the average current, the expectation value of the operator in
Eq.~(\ref{CurrentOperator}) can be expressed as
\begin{eqnarray}
I(z_1,t_1) = \frac{e}{8m^*}
\lim_{x_2\rightarrow x_1}
\int d^2 r_{1\|}
\left(\partial_{z1}-\partial_{z2}\right)\qquad&\nonumber\\
\mbox{\mbox{Tr}}\left\{{\hat G}^K(x_1,x_2)
-\hat\tau_3\left[{\hat G}^R(x_1,x_2)-{\hat G}^A(x_1,x_2)\right]
\right\}
\,,&
\label{AverageCurrent}
\end{eqnarray}
for the current flowing through the junction
along the $z$-axis. When we insert the quasiclassical envelope expansions,
Eqs.~(\ref{parallel_momentum})-(\ref{Zaitsev_expansion}), the
derivatives produce a factor
$ip_{fz}(\nu+\mu)=2ip_{fz}\delta_{\nu\mu}$, and the current takes the
form
\begin{equation}\begin{split}
I(z_1,t) &= \Area\frac{ie}{2} \lim_{z_2\rightarrow z_1}
\int \frac{d^2p_{\|}}{(2\pi)^2} \int \frac{d\epsilon}{2\pi}
\sum_{\nu}\,\nu\,\,\times\\
&\quad \mbox{\mbox{Tr}}\left\{
\tau_3{\hat C}^K_{\nu\nu}(z_1,z_2,{\bf p}_{\|};\epsilon,t)
\right\},
\end{split}\end{equation}
where we neglect terms where the derivative act on quasiclassical
Green's functions since they are down by a factor $(p_f\xi_0)^{-1}$
compared to the leading term above. Also note that the term $\hat C^R-\hat
C^A$ drops out because spectral current density is odd in energy to
this order in $(p_f\xi_0)^{-1}$.

When we insert the quasiclassical expressions,
Eqs.~(\ref{parallel_momentum})-(\ref{Zaitsev_expansion}), into the
expression for the noise, Eq.~(\ref{ExchangeNoise}), the derivatives
produce a factor $(ip_{fz})^2(\tilde\nu+\mu)(\nu+\tilde\mu)
=(2ip_{fz})^2\delta_{\tilde\nu\mu}\delta_{\tilde\mu\nu}$, where the
indices without (with) a tilde belong to the first (second) Green's
function. Thus, to quasiclassical accuracy we obtain,
\begin{eqnarray}\label{KhlusNoise}
&\hspace*{-6mm}S(z_1,t_1,{\tilde t}_1) =
\Area \displaystyle{\frac{e^2}{2}\lim_{\tilde
z_1\rightarrow z_1}} \int \frac{d^2p_{\|}}{(2\pi)^2}
\,\sum_{\nu\mu}\nu\mu\,\,\times \,\; & \\
&\mbox{Tr}\{\hat C_{\nu\mu}^<(\tilde z_1,z_1,\vp_{\|};\tilde{t}_1,t_1)\hat\tau_3
            \hat C_{\mu\nu}^>(z_1,\tilde z_1,\vp_{\|};t_1,\tilde{t}_1)\hat\tau_3 \}
\,,
&\nonumber
\end{eqnarray}
where $\Area$ is the cross-sectional area of the junction - which is
the result previously obtained by Khlus\cite{khl87} for conventional NS junctions.

The spectral density of the noise is given by the Fourier
transform,\cite{but00} $S(z,t,\omega) = \int d\tau e^{i\omega \tau}
S(z,t,\tau)$, which for general non-equilibrium conditions depends on
time. Here we consider the d.c. limit for the voltage-biased
junction. This steady-state limit is independent of time, so we drop
the time argument from here on.

The drones, as well as the quasiclassical correlation functions,
$\hat C_{\nu\nu}^{\gtrless}(\tilde z_1,z_1)=\hat C^K \pm (\hat
C^R-\hat C^A)$, are continuous and single-valued for $\tilde
z_1=z_1$.\cite{mil88} Thus, we may take the limit $\tilde
z_1\rightarrow z_1$; the diagonal functions are proportional to the
corresponding quasiclassical Green's functions,
\begin{equation}
\lim_{z_2\rightarrow z_1}\hat C_{\nu\nu}^{\gtrless}(z_1,z_2) =
\frac{\hg_{\nu}^{\gtrless}(z_1)}{2\pi} \,,
\end{equation}
and we introduce the following notation for the drones,
\begin{equation}
\lim_{z_2\rightarrow z_1}\hat C^{\gtrless}_{\nu(-\nu)}(z_1,z_2) = \frac{\hat
d^{\gtrless}_{\nu(-\nu)}(z_1)}{2\pi} \,.
\end{equation}
Thus, the current is written as
\begin{equation}\label{current}
I = -eN_f\Area\int\frac{d\epsilon}{4\pi i} \sum_{\nu=\pm 1}\,\nu\,\,
\mbox{Tr}\left\langle v_{fz}\hat\tau_3\hg^K_{\nu}\right\rangle_{\nu}
\,,
\end{equation}
which is the $z$-component of the more general quasiclassical result for the
current density,
\begin{equation}\label{current_full}
\vj = -eN_f\int\frac{d\epsilon}{4\pi i}
\mbox{Tr}\left\langle\vv_f\hat\tau_3\hg^K\right\rangle_{\vp_f}
\,,
\end{equation}
where $N_f$ is the normal-state density of states at the Fermi level and
$\langle\ldots\rangle_{\vp_f}=N_f^{-1}\int\frac{d^2p_f}{((2\pi)^3|\vv(\vp_f)|}(\ldots)$
denotes a Fermi surface average.

Similarly, the local noise spectrum may be written as
\begin{eqnarray}\label{noise}
S = e^2 N_f\Area
    \int\frac{d\epsilon}{16\pi^2}\sum_{\nu=\pm 1}\,\times\qquad & & \\
\mbox{Tr}  \left\langle v_{fz}
    \left(\hat g^<_{\nu}\hat\tau_3
          \hat g^>_{\nu}\hat\tau_3  -
          \hat d^<_{\nu(-\nu)}\hat\tau_3
          \hat d^>_{(-\nu)\nu}\hat\tau_3
    \right)\right\rangle_{\nu}
\,,&&\nonumber
\end{eqnarray}
where $\langle\ldots\rangle_{\nu}$ denotes a Fermi surface average
restricted to $\mbox{sgn}({\bf p}_f\cdot\hat z)=\nu$. Although the
noise formula in Eq. ({\ref{noise}) depends on the drones, they do not
require independent calculation from their equations of motion, but
are expressed in terms of the quasiclassical propagators (see
below). We note that in writing down Eq.~(\ref{noise}), the limit
$\omega\rightarrow 0$ was taken.

\subsection{Interface Boundary Conditions}

In order to compute the transport current and noise spectrum we must solve transport
equations for the quasiclassical propagators, and drones, with appropriate boundary
conditions describing the junction. Quasiclassical boundary conditions describing partially
transmitting interfaces between conducting electrodes were derived by Zaitsev\cite{zai84} and
Kieselmann\cite{kie87} for non-magnetic junctions, and by Millis et al.\cite{mil88} for
magnetically active interfaces. Zaitsev and Kieselmann's boundary conditions are a set of
nonlinear equations connecting the quasiclassical propagators for incoming and outgoing
trajectories at the interface. The material parameters entering these boundary conditions are
the reflection ($\cal R$) and transmission (${\cal D}=1-{\cal R}$) probabilities for
quasiparticles when both electrodes are in their normal-states. This
formulation is valid for arbitrary transparency.

\begin{figure}[t]
\includegraphics[width=8cm]{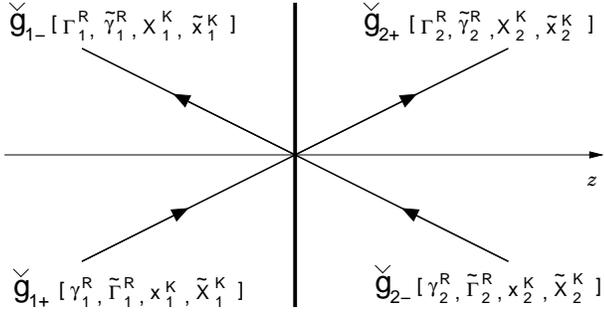}
\caption{\small
We label Green's functions by an index $1$ for the left (normal) electrode
and $2$ for the right (superconducting) electrode, and a direction index $\pm$
which denotes the sign of the projection of the Fermi momentum $\vp_f$ on
the $z$-axis. The arrows on each line indicates the direction of $\vp_f$.
Also shown is the notation for the propagators, coherence amplitudes and
distribution functions.
}
\label{junction}
\end{figure}

The quasiclassical boundary conditions, when one or both electrodes are superconducting,
incorporate the effects of particle-hole coherence of the excitations in the superconducting
electrodes, \textsl{and} the presence of additional channels for reflection and transmission
via branch conversion scattering between particle- and
hole-like branches of excitations.

A powerful method for handling the interplay between these coherence effects and
interface scattering was provided by Eschrig's reduction\cite{esc00} of Zaitsev and
Kieselmann's boundary conditions using Schelankov's projection operators\cite{she85}
and the Ricatti parametrization for the quasiclassical propagators.\cite{nag93,sch95,esc99}
The boundary condition is expressed in terms of coherence functions, $\gamma^R$ and
$\tilde\gamma^R$, and distribution functions, $x^K$ and $\tilde x^K$. The coherence
functions have a natural interpretation as local amplitudes for branch conversion;
$\gamma^R$ for $h\rightarrow e$ and $\tilde\gamma^R$ for $e\rightarrow h$. Below we
express these boundary conditions in terms of generalized scattering amplitudes, which in
the clean limit are directly related to well known scattering amplitudes found in
scattering theory.\cite{but00} The considerations below are valid for general
non-equilibrium situations, but the results presented below are limited to
time-independent states. For non-stationary states, all multiplications are replaced by
time-convolutions, which in general prevents analytic computations.

We adopt the notation used in Ref. \onlinecite{esc00} for the coherence amplitudes and
distribution functions. The labelling for functions defined on incoming and outgoing
trajectories is also indicated in Fig. \ref{junction}.
For the distribution functions, the boundary conditions can be written as
\begin{eqnarray}
X_1^K        &= R^R_{ee}x_1^K + \bar T^R_{ee}x_2^K
                + (-\bar T^R_{eh})\tilde x_2^K,\label{x_bc1}\\
\tilde X_1^K &= R^R_{hh}\tilde x_1^K + (-\bar T^R_{he})x_2^K
                + \bar T^R_{hh}\tilde x_2^K,\label{x_bc2}\\
X_2^K        &= T^R_{ee}x_1^K + (-T^R_{eh})\tilde x_1^K
                + \bar R^R_{ee}x_2^K,\label{x_bc3}\\
\tilde X_2^K &= (-T^R_{he})x_1^K + T^R_{hh}\tilde x_1^K
                + \bar R^R_{hh}\tilde x_2^K,\label{x_bc4}
\end{eqnarray}
where the scattering amplitudes are defined as
\begin{align}\label{scattering_amplitudes-b}
r^R_{ee}      &= R_{1l}^R/r, & \bar t^R_{ee} &= D_{1l}^R/d,
              & \bar t^R_{eh} &= rdA_{1l}^R,\\
r^R_{hh}      &= \tilde R_{1l}^R/r, & \bar t^R_{hh} &= \tilde D_{1l}^R/d,
              & \bar t^R_{he} &= rd\tilde A_{1l}^R,\\
\bar r^R_{ee} &= R_{2l}^R/r, & t^R_{ee} &= D_{2l}^R/d,
              & t^R_{eh} &= rdA_{2l}^R,\\
\bar r^R_{hh} &= \tilde R_{2l}^R/r, & t^R_{hh} &= \tilde D_{2l}^R/d,
              & t^R_{he} &= rd\tilde A_{2l}^R\,.\label{scattering_amplitudes-e}
\end{align}
The right-hand sides of Eqs. (\ref{scattering_amplitudes-b}-\ref{scattering_amplitudes-e}) are defined in
Eqs. (D1-D5) of Ref.~\onlinecite{esc00}. Note
that all quantities depend on trajectory angle ${\bf p}_f$ and energy $\epsilon$,
but not on spatial coordinates ${\bf R}$ since they are evaluated at the junction.
The normal-state tunnel barrier transmission and reflection amplitudes are denoted $d$
and $r$, respectively, while the corresponding probabilities are denoted by $\mathcal{D}=d^2$
and $\mathcal{R}=r^2$.\footnote{The phases of $r$ and $d$ does not play a role in our
calculations and can be eliminated; thus, both $r$ and $d$ are taken to be real.}

The effective transmission and reflection amplitudes, including Andreev scattering, are
denoted by $t^R_{\alpha\beta}$ and $r^R_{\alpha\beta}$, while the corresponding probabilities
are denoted as $T^R_{\alpha\beta}$ and $R^R_{\alpha\beta}$. For example, $r^R_{hh}$ is the
amplitude for reflection of a hole on the left side of the junction, while $\bar r^R_{ee}$ is
the amplitude for reflection of an electron on the right side. Similarly, $\bar t^R_{he}$ is
the transmission amplitude for an electron from the right side to the left side, including
branch conversion into a hole. All quantities with a bar refer to excitations
originating from the right electrode.

The remaining amplitudes are the Andreev reflections, which appear via the boundary
conditions for the coherence functions,
%
\begin{eqnarray}\label{scattering_amplitudes2}
r^R_{he}      =& \tilde \Gamma^R_1
             &= r^R_{hh}\tilde\gamma^R_1r + \bar t^R_{hh}\tilde\gamma^R_2d,\\
r^R_{eh}      =& \Gamma^R_1
             &= r^R_{ee}\gamma^R_1r + \bar t^R_{ee}\gamma^R_2d,\\
\bar r^R_{he} =& \tilde\Gamma^R_2
             &= \bar r^R_{hh}\tilde\gamma^R_2r + t^R_{hh}\tilde\gamma^R_1d,\\
\bar r^R_{eh} =& \Gamma^R_2
             &= \bar r^R_{ee}\gamma^R_2r + t^R_{ee}\gamma^R_1d
\,.
\end{eqnarray}

In the Appendix we summarize the results
for the Andreev reflection probabilities and scattering probabilities that
enter the Keldysh distribution functions in Tables (\ref{ARamplitudes}-\ref{TRprobabilities}).
These probabilities are expressed in terms of the normal-state barrier transmission
and reflection probabilities, and the coherence amplitudes for particle and
hole excitations.

We introduce the notation,
\begin{equation}
  |\alpha\rangle = \left(
  \begin{array}{c}
    1\\
    -i\sigma_y\alpha
  \end{array}
  \right),\hspace{1cm}
  \langle\alpha| = \left(1,\;\; -i\sigma_y\alpha^* \right)
\,,
\end{equation}
which is convenient for evaluating observables. For example, to calculate
the charge current we need
\begin{equation}\label{Tracerules}
\mbox{Tr}\left\{ \hat\tau_3 |\alpha\rangle \langle\beta| \right\}
  = 2(1+\alpha\beta^*)
\,,
\end{equation}
where the factor $2$ comes from the spin trace.
The Keldysh Green's functions at the junction
can now be written in a rather compact form,
\begin{equation}\label{gK_scattering}\begin{split}
    \hat g^K_{1+} &= -2\pi i N_1^{-1}
    \left[ x_1^K |r^R_{he}\rangle \langle r^R_{he}|
      + \tilde X_1^K \hat\tau_1 |\gamma^R_1\rangle
      \langle\gamma^R_1|\hat\tau_1 \right],\\
    \hat g^K_{1-} &= -2\pi i N_2^{-1}
    \left[ \tilde x_1^K \hat\tau_1|r^R_{eh}\rangle \langle r^R_{eh}|\hat\tau_1
      + X_1^K |\tilde\gamma^R_1\rangle \langle \tilde\gamma^R_1| \right],\\
    \hat g^K_{2-}  &= -2\pi i N_3^{-1}
    \left[ x_2^K |\bar r^R_{he}\rangle \langle\bar r^R_{he}|
      + \tilde X_2^K \hat\tau_1|\gamma^R_2\rangle
      \langle\gamma^R_2|\hat\tau_1 \right],\\
    \hat g^K_{2+} &= -2\pi i N_4^{-1}
    \left[ \tilde x_2^K \hat\tau_1 |\bar r^R_{eh}\rangle
      \langle\bar r^R_{eh}|\hat\tau_1
      + X_2^K |\tilde\gamma^R_2\rangle \langle \tilde\gamma^R_2| \right]
\,,
\end{split}\end{equation}
where we introduced the denominators, $N_i=|\zeta_i|^2$ for $i=1..4$ with
\begin{align}
  \zeta_1 &= 1+\gamma^R_1 r^R_{he}
  & \zeta_3 &= 1+\gamma^R_2 \bar r^R_{he},\\
  \zeta_2 &= 1+\tilde\gamma^R_1 r^R_{eh}
  & \zeta_4 &= 1+\tilde\gamma^R_2 \bar r^R_{eh}
\,.
\end{align}
Note that all denominators of the scattering probabilities in
Eqs.~(\ref{x_bc1})-(\ref{x_bc4}) (see Table \ref{TRprobabilities} in the Appendix)
are cancelled by the denominators, $N_i$, in Eqs.~(\ref{gK_scattering}); e.g.
$R^R_{ee}N_1^{-1}=(A/\zeta_1)*(\zeta_1/|Z|^2)=A/|Z|^2=R_{ee}^{R'}$.
As a consequence, a common
denominator, $|Z|^2$, enters all Keldysh propagators,
\begin{equation}\begin{split}
Z &= 1 + R(\gamma^R_2\tilde\gamma^R_2+\gamma^R_1\tilde\gamma^R_1)\\
  &\quad + D(\gamma^R_2\tilde\gamma^R_1+\tilde\gamma^R_2\gamma^R_1)
  + \gamma^R_1\tilde\gamma^R_1\gamma^R_2\tilde\gamma^R_2
\,.
\end{split}\end{equation}
This function also appears as the denominator of the retarded Green's function.
Thus, the zeroes of $Z$ determine the local spectrum of excitations, including
interface bound states, at the junction.

We note that the scattering amplitudes defined above do not exactly
coincide with the ones obtained in scattering theory. There are
missing pre-factors, which are hidden in the matrices
$|\alpha\rangle\langle\alpha|$ in Eqs.~(\ref{gK_scattering}), and in
the distribution functions $x^K$ and $\tilde x^K$ [e.g. in
equilibrium, $x^K=(1-|\gamma^R|^2)\tanh(\epsilon/2T)$]. Inspection
shows that our generalized scattering amplitudes can be interpreted
as describing the scattering of \textsl{locally} defined excitations at the
junction, while the factors coming from the matrices and distribution
functions gives a spectral renormalization due to Andreev reflection
along the trajectories leading up to (and away from) the interface.
For example, when a charge current is computed, these
renormalizations can be absorbed into the scattering amplitudes,
which then coincide with results from scattering theory. However, we
retained the above definitions since they appear naturally in the
boundary condition for Green's functions.

The above considerations are applicable to stationary states of two
coupled superconductors driven out of equilibrium. For the special
case in which the left electrode, is in the normal state,
$\gamma^R_1=\tilde\gamma^R_1=0$. Using Eq.~(\ref{Tracerules}) we
obtain the current computed at the junction on the left side of the
barrier
\begin{equation}\label{NIS_current}
\begin{split}
I &= eN_f\Area \int d\epsilon \left\langle v_{fz} \left[
x_1^K(1+R^R_{he}-R^R_{ee})\right.\right.\\
&\quad + \tilde x_1^K(1+R^R_{eh}-R^R_{hh})\\
&\quad + x_2^K(\bar T^R_{he} - \bar T^R_{ee}) +
\left.\left.\tilde x_2^K(\bar T^R_{eh} - \bar T^R_{hh})
\right] \right\rangle_{{\bf p}_f}.
\end{split}
\end{equation}
Explicit expressions for the effective scattering amplitudes are given in
Table~\ref{S-table} for the NIS junction. Equation (\ref{NIS_current}) is
valid for arbitrary stationary non-equilibrium situations, including spatially
dependent coherence and distribution functions. Current conservation is
guaranteed for self-consistent calculations.

\begin{table}
\caption{\small Scattering amplitudes at an NIS juntion. The common denominator is
$Z=1+\mathcal{R}\gamma^R\tilde\gamma^R$}
\begin{tabular}{|l|l|l|l|}
\hline $r^R_{ee} = \frac{r(1+\gamma^R\tilde\gamma^R)}{Z}$ & $r^R_{hh} = \frac{r(1+\gamma^R\tilde\gamma^R)}{Z}$
& $\bar r^R_{ee} = \frac{r}{Z}$ &
$\bar r^R_{hh} = \frac{r}{Z}$\\
\hline $r^R_{he} = \frac{\mathcal{D}\tilde\gamma^R}{Z}$ & $r^R_{eh} = \frac{\mathcal{D}\gamma^R}{Z}$ & $\bar
r^R_{he} = \mathcal{R}\tilde\gamma^R$ &
$\bar r^R_{eh} = \mathcal{R}\gamma^R$\\
\hline $t^R_{ee} = \frac{d}{Z}$ & $t^R_{hh} = \frac{d}{Z}$ & $\bar t^R_{ee} = \frac{d}{Z}$ &
$\bar t^R_{hh} = \frac{d}{Z}$\\
\hline $t^R_{he} = \frac{rd\tilde\gamma^R}{Z}$ & $t^R_{eh} = \frac{rd\gamma^R}{Z}$ & $\bar t^R_{he} =
-\frac{rd\tilde\gamma^R}{Z}$ &
$\bar t^R_{eh} = -\frac{rd\gamma^R}{Z}$\\
\hline
\end{tabular}\label{S-table}
\end{table}

\subsection{Asymptotic Boundary Conditions}

In the reservoir regions, far from the junction, the distribution functions take the equilibrium forms,
shifted by the local potential,
\begin{equation}\begin{split}
F_1(x\rightarrow -\infty,\epsilon) &= \tanh[(\epsilon-eV)/(2T)],\\
F_2(x\rightarrow +\infty,\epsilon) &= \tanh[\epsilon/(2T)].
\end{split}\end{equation}
The hole distributions follow by symmetry $\tilde F(\epsilon) = F(-\epsilon)$. We  neglect
processes where quasiparticles scattered at the junction are scattered back and impinge on
the junction before they equilibrate. The above distribution functions then
serve as incoming distribution functions in the boundary condition at the junction. We
shall also assume that the transparency of the junction is sufficiently small,
$\mathcal{D}\ll 1$, that the current flowing throught the system (which is proportional to
$\mathcal{D}$) due to the applied voltage is small. Then, to lowest order in $\mathcal{D}$ we
can neglect the effect of the current on the order parameter and write
\begin{equation}
f^K({\bf p}_f,{\bf R};\epsilon) = \left[ f^R - f^A \right] \tanh\frac{\epsilon}{2T}
\,,
\end{equation}
which is the local equilibrium form for the off-diagonal Keldysh propagator. We
note that these assumptions will be valid also for high-transparency point
contacts and for wide junctions with transport primarily through a
high-transparency pinhole, since the current in those cases are reduced by the
small conducting area $A\ll\pi\xi_0^2$, where $\xi_0=v_f/T_c$ is the
superconducting coherence length.

Under these assumptions, the interface distribution
functions are,
\begin{eqnarray}
x^K_1=F_1\,,\quad & x^K_2=(1-|\gamma^R|^2)F_2&
\\
\tilde x^K_1=\tilde F_1\,,\quad & \;\;\tilde x^K_2=-(1-|\tilde\gamma^R|^2)F_2&
\,,
\end{eqnarray}
where we drop the subscript $2$ on coherence functions since they are
superfluous for an NIS system. The $x_2^K$ and $\tilde{x}_2^K$ terms then cancel in
Eq. (\ref{NIS_current}) by the general tilde-symmetry, which relates any quantity
$\tilde q$
to its partner $q$ as
$ \tilde q({\bf p}_f,{\bf R};\epsilon,t)=q(-{\bf p}_f,{\bf R};-\epsilon,t)^*$.
However, the $x_1^K$ and $\tilde{x}_1^K$ terms cancel only at zero bias
because particles and holes have opposite charge.

\subsection{Drone Green's functions and noise}

To compute the noise in Eq.~(\ref{noise}) we need to also compute the drone amplitudes
$\check d_{\nu(-\nu)}$. The relations connecting the drones to the quasiclassical propagators
are the same equations used to obtain the nonlinear boundary condition connecting the
quasiclassical propagators $\check g_{1\pm}$ and $\check g_{2\pm}$.\cite{zai84,mil88}
Thus, we define the symmetric combination of Green's functions on the two sides
($i=1,2$) of the interface as
\begin{equation}
\check g_{is} = \check g_{i+} - \check g_{i-},
\end{equation}
and symmetric and anti-symmetric combinations of drones as
\begin{equation}\begin{split}
\check d_{is} &= \check d_{i+-} + \check d_{i-+},\\
\check d_{ia} &= \check d_{i+-} - \check d_{i-+}.
\end{split}\end{equation}
The necessary relations are then
\begin{equation}\begin{split}
\check d_{1s} &= \frac{1}{2\sqrt{\cal R}}
\left[(1+{\cal R})\check g_{1s} - {\cal D}\check g_{2s}\right],\\
\check d_{2s} &= \frac{1}{2\sqrt{\cal R}}
\left[{\cal D}\check g_{1s}-(1+{\cal R})\check g_{2s}\right],\\
4\pi i \check d_{1a} &=\check g_{1s}\check d_{1s}-\check g_{2s}\check d_{2s},
\end{split}\end{equation}
where the first two relations come from the boundary condition, and the last relation
is derived by making use of the normalization condition for Green's functions [c.f.
Eqs.~(29)-(30) of Ref.~\onlinecite{mil88}].
We note that we are content with solving for the drones on the left side.
Explicit expressions of the drones can then be written down by using the Green's
function $\check g$ written in terms of scattering amplitudes in Table~\ref{S-table}. The
retarded and advanced drones are
%
\begin{align}
\hat d_{1s}^R &= -2\pi i r^R_{ee} \hat\tau_3,
& \hat d_{1s}^A &= +2\pi i r^{R*}_{ee} \hat\tau_3,\\
\hat d_{1a}^R &= +2\pi i r^R_{ee} \hat 1, & \hat d_{1a}^A &= +2\pi i r^{R*}_{ee}
\hat 1\,,
\end{align}
while Keldysh drones take the form
\begin{equation}
\begin{split}
B &=x_1^K r^R_{ee}r^{R*}_{he}+
\tilde x_1^K r^{R*}_{hh}r^R_{eh}-
x_2^K\bar t^R_{ee}\bar t^{R*}_{he} -
\tilde x_2^K\bar t^{R*}_{hh}\bar t^R_{eh}\,,\\
\hat d_{1s}^K &= -2\pi i \begin{pmatrix} x_1^K (r^R_{ee}+r^{R*}_{ee}) &
-i\sigma_y B\\
-i\sigma_y B^* & \tilde x_1^K (r^R_{hh}+r^{R*}_{hh})
\end{pmatrix},\\
\hat d_{1a}^K &= +2\pi i \begin{pmatrix} x_1^K (r^R_{ee}-r^{R*}_{ee}) &
-i\sigma_y B\\
+i\sigma_y B^* & -\tilde x_1^K (r^R_{hh}-r^{R*}_{hh})
\end{pmatrix}\,.
\end{split}
\end{equation}
where $i\sigma_y$ is the Pauli matrix that describes spin-singlet pairing.

Since the noise is expressed in terms of
$\hg^{\gtrless}=\hg^K\pm (\hg^R-\hg^A)$ and
$\hat d^{\gtrless}=\hat d^K\pm (\hat d^R-\hat d^A)$,
we get terms from the Keldysh parts which depend explicitly on the distribution
functions, and purely spectral
terms that do not contain any distribution functions. Thus, we
separate the noise into two terms, $S=S^{R-A}+S^K$, where
\begin{eqnarray}\label{NIS_noise_RA}
S^{R-A}({\bf p}_f,x=0^-;\epsilon) =
4 \left[ (1+R^R_{he}-R^R_{ee}) + \right. \nonumber \\
  \left.(1+R^R_{eh}-R^R_{hh}) \right]
\,,
\end{eqnarray}

\begin{widetext}
\begin{equation}\label{NIS_noise}\begin{split}
S^K({\bf p}_f,x=0^-;\epsilon) &=
- (x_1^K)^2 2 \left[ 1+R^R_{he}-R^R_{ee} \right]^2
- (\tilde x_1^K)^2 2 \left[ 1+R^R_{eh}-R^R_{hh} \right]^2\\
&\quad - \left(x_2^K\right)^2 2(\bar T^R_{he}-\bar T^R_{ee})^2
- \left(\tilde x_2^K\right)^2 2(\bar T^R_{eh}-\bar T^R_{hh})^2\\
&\quad + x_1^K \tilde x_1^K 4|r^R_{he}r^{R*}_{hh}+r^R_{ee}r^{R*}_{eh}|^2
- x_1^K x_2^K 4|r^R_{he}\bar t^{R*}_{he}+r^R_{ee}\bar t^{R*}_{ee}|^2
+ x_1^K \tilde x_2^K 4|r^R_{he}\bar t^{R*}_{hh}-r^R_{ee}\bar
t^{R*}_{eh}|^2\\
&\quad + \tilde x_1^K x_2^K 4|r^R_{eh}\bar t^{R*}_{ee}-r^R_{hh}\bar
t^{R*}_{he}|^2
- \tilde x_1^K \tilde x_2^K 4|r^R_{eh}\bar t^{R*}_{eh}+r^R_{hh}\bar
t^{R*}_{hh}|^2
+ x_2^K\tilde x_2^K 4|\bar t^R_{he}\bar t^{R*}_{hh}+\bar t^R_{ee}\bar
t^{R*}_{eh}|^2
\,.
\end{split}\end{equation}
\end{widetext}
The above results are valid for general non-equilibrium distribution.
The distribution functions $x_i^K$ and $\tilde{x}_i^K$ can always be expressed
as local equilibrium distributions plus `anomalous' non-equilibrium distributions.
Then the purely spectral terms, $S^{R-A}$, are cancelled exactly by local equilibrium
terms in $S^K$ that do not contain a Fermi function.

Equations (\ref{NIS_current}) and (\ref{NIS_noise}), combined with Tables (\ref{S-table})
for the reflection and transmission probabilities are the central equations
needed for calculating the conductance and noise spectrum for  NIS
junctions with disorder, unconventional pairing and interface screening currents.
These formulas are expressed in a form that is closely related to the wave-function-based
scattering theory applicable to clean systems. This connection is based on the
identification between the scattering amplitudes in the wave function approach
and the retarded Ricatti amplitude, $\gamma^R$, which, in the clean limit,
reduces to the local Andreev reflection amplitude, $v/u$. However, the Ricatti representation
for the propagators is more general, and is capable of incorporating the effects of
disorder and inelastic scattering. In our formulation, all observables
can then be expressed in terms of the generalized scattering amplitudes collected in
Table~\ref{S-table}, and in the tables in the Appendix. However, the generalized
scattering amplitudes are only defined in terms of the quasiclassical Green's functions,
through the Ricatti parametrization. We emphasize this fact by keeping the
superscript $R$ on all quantities defined in terms of the retarded Green's function.

In summary, to compute the conductance and noise spectrum in voltage-biased NIS
junctions we solve the quasiclassical transport equations (\ref{transport_eq})
for $\check g$ self-consistently with the gap equations (\ref{gap_eqs}), the
$t$-matrix equations (\ref{t-matrix_RA}-\ref{t-matrix_K}), and the surface
coupling to the screening currents, (Eqs. \ref{Doppler}-\ref{ps}). We then
compute the effective reflection and transmission probabilities, and distribution
functions and use Eqs. (\ref{NIS_current}) and (\ref{NIS_noise})
to calculate the conductance and the noise spectrum.

\section{Conductance and Differential Shot Noise}\label{SecNoise}

In the following we use these results to calculate the conductance and noise spectrum for NIS junctions with
d-wave superconductors. In the zero-temperature limit, Eq. (\ref{NIS_current}) for the current can be written
as
\begin{equation}\label{current_T0}
\begin{split}
eR_nI(V) & = eV + \frac{1}{\mathcal{D}}\int_{-eV}^{0}d\epsilon\left\langle v_{fz} \left[{\cal R}(\vp_f)
\right.\right.
\\
&\quad + \left.\left. R_{he}^R(\epsilon,\vp_f)-R_{ee}^R(\epsilon,\vp_f)\right]\right\rangle_{\vp_f\cdot\hat
z>0} \,.
\end{split}
\end{equation}
The corresponding zero-temperature shot noise, computed at $z=0^-$, from Eqs. (\ref{NIS_noise}) and Table
(\ref{S-table}), takes the form,
\begin{equation}\label{noise_T0}
\begin{split}
R_nS(V) & = \frac{2}{\mathcal{D}}\int_{-eV}^0 d\epsilon\left\langle v_{fz}
    \left\{R_{ee}^R(\epsilon,\vp_f)[1-R_{ee}^R(\epsilon,\vp_f)]
\right.\right.
\\
&    \quad+R_{he}^R(\epsilon,\vp_f)[1-R_{he}^R(\epsilon,\vp_f)]
\\
&    \quad+\left.\left.2R_{ee}^R(\epsilon,\vp_f)R_{he}^R(\epsilon,\vp_f)\right\} \right\rangle_{\vp_f\cdot\hat
z>0} \,.
\end{split}
\end{equation}
The normal-state junction resistance is given by $R_n^{-1}=2\Area e^2 N_f v_f\mathcal{D}$, where
$\mathcal{D}\equiv\langle\cos\theta\mathcal{D}(\vp_f)\rangle_{\vp_f\cdot\hat z>0}$ is the {\sl transport}
barrier transparency; $\cos\theta=\hat p_f\cdot\hat z\ge 0$ is the angle of incidence measured relative to the
$z$-axis. Note that in the normal-state limit for the superconducting electrode, the Andreev reflection
probability vanishes, $R_{he}^R(\epsilon,\vp_f)\rightarrow 0$, the effective $e\rightarrow e$ reflection
probability reduces to $R_{ee}^R(\epsilon,\vp_f)\rightarrow {\cal R}(\vp_f)$, and the integrand of
Eq.~(\ref{current_T0}) vanishes. Thus, we recover Ohm's law for the junction I-V characteristic. Similarly,
the shot noise in the NIN limit is proportional to $\langle v_{fz}\mathcal{R}(\vp_f)\mathcal{D}(\vp_f)
\rangle_{\vp_f}$, which for small transparency corresponds to the Schottky result $S=2eI$. For higher
transparency the current noise is reduced compared to the Schottky result.

The above expressions have in the clean limit the same forms as well known scattering theory
results.\cite{blo82,tan95,but00,ana96,hes96,zhu99,tan00,lof02a} Eq.~(\ref{current_T0}) also agrees with
calculations of spectral current densities including impurity scattering and subdominant pairing, in
Ref.~\onlinecite{esc00}.

In the following we present calculations of the zero temperature conductance $\partial I/\partial V$ and
differential shot noise, $\partial S/\partial V$, and focus on effects of magnetic fields, impurity
scattering, and subdominant pairing. The exact angle dependence of the tunnelling probability is not
particularly important for our purposes; so we take it to have the form predicted by an interface
$\delta$-function potential,
\begin{equation}
\mathcal{D}(\vp_f) = \mathcal{D}_0 \frac{\cos^2\theta}{1-\mathcal{D}_0\sin^2\theta} \,,
\end{equation}
where $\mathcal{D}_0$ is the transparency for normal incidence.


\subsection{Pure $d_{x^2-y^2}$-wave: effects of a magnetic field}

\begin{figure}[t]
\includegraphics[width=8cm]{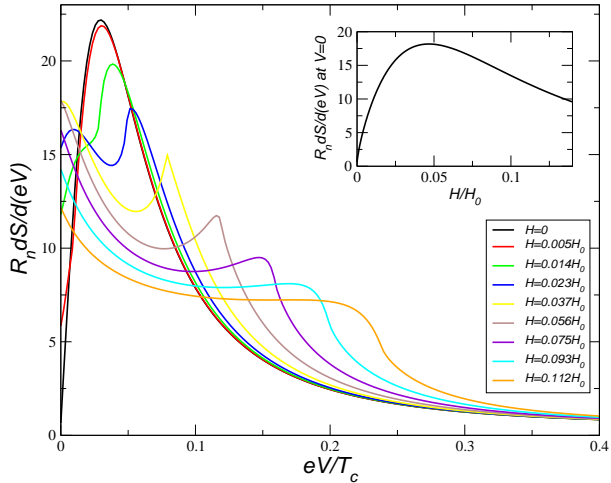}
\caption{\small Zero temperature differential shot noise as a function
 of voltage in the clean limit for several different external magnetic
 fields.  Inset: The field evolution of the zero-voltage differential
 shot noise. The transparency of the interface is
 $\mathcal{D}_0=0.1$.}\label{dSdV_Hext}
\end{figure}

In zero external magnetic field, the angle resolved differential shot noise is
suppressed to zero at zero voltage and has a peak at
$\sim\mathcal{D}(\vp_f)|\Delta_0(\vp_f)|$, where $\Delta_0(\vp_f)=
\Delta_{B_{1g}}(z\rightarrow\infty)\eta_{B_{1g}}(\vp_f)$ is the gap in the
bulk. This result is due to the resonant enhancement of Andreev reflection by the
surface bound state: around the bound state energy ($\epsilon=0$) within an energy
interval set by the bound state width $w_b(\vp_f) = a\mathcal{D}(\vp_f)|\Delta_0(\vp_f)|$, the probability of Andreev reflection is enhanced to
unity $R_{he}^R(\epsilon=0,\vp_f) = 1$ independently of the smallness of the
transparency and independently of the shape of the order parameter near the
junction. The numerical prefactor $a$ is due to the reduction of the bound state
width caused by the suppression of the order parameter near the surface. It was
computed for small $\mathcal D$ in Ref.~\onlinecite{bar00b} and can be
estimated to be approximately $1/4$. As the Andreev reflection probability is
enhanced to unity, the normal reflection probability is reduced to zero,
$R_{ee}^R(\epsilon=0,\vp_f) = 0 $. The result of zero noise at $V=0$ then
follows directly from Eq.~(\ref{noise_T0}). The suppression of $\partial S/\partial
V$ to zero at zero voltage for zero field is robust under angle integration since
the zero-energy bound state is dispersionless. The satellite peak will be located at
a voltage of the order $ \langle w_b(\vp_f) \rangle_{\vp_f} \approx
{\mathcal D}_0 T_c / 2\pi $. This noise-\textsl{less} character of the zero-energy
bound states in a clean system was recently discussed in
Refs.~\onlinecite{zhu99,tan00,lof02a}.

In an externally applied magnetic field, the screening currents produce a Doppler
shift of the spectrum. The bound state resonance is shifted accordingly. The point of
suppressed noise is then shifted to finite voltage and the peak in $\partial
S/\partial V$ is pushed to higher voltages linearly with increasing magnetic field.
These characteristics of the field evolution of the shot noise spectrum are shown in
Fig.~\ref{dSdV_Hext}. The dispersion of the Doppler-shifted ABS leads to non-zero
differential shot noise at all voltages. In particular, at zero voltage the
differential shot noise develops with increasing magnetic field strength as shown in
the inset of Fig.~\ref{dSdV_Hext}.

\subsection{$d_{x^2-y^2}$ pairing with impurity scattering:
             Andreev vs. tunnel limits}

The sensitivity of the noise to changes in the low-energy surface
excitation spectrum implies that the results for $S(V)$ in clean
d-wave superconductors\cite{zhu99,tan00,lof02a} are strongly
modified by disorder. Here we consider the effects of impurity
scattering on the noise spectrum.

\begin{figure}[t]
\includegraphics[width=8cm]{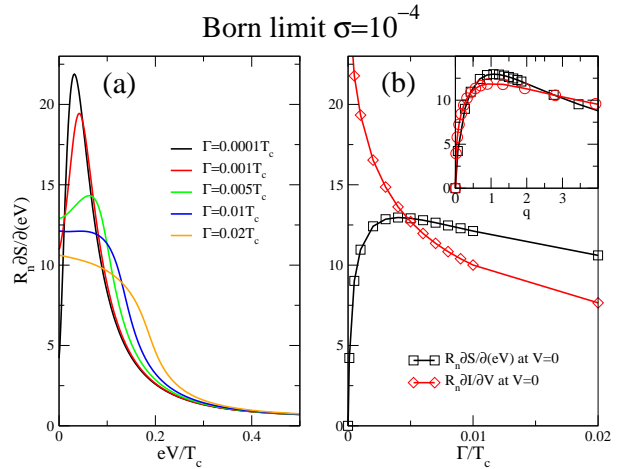}
\caption{\small (a) Differential shot noise for several different pair breaking
parameters $\Gamma$ for scattering in the Born limit ($\sigma=10^{-4}$). (b) The
zero-voltage value of differential shot noise and conductance as a function of
$\Gamma$. The squares and diamonds are the numerically computed results, while the
lines are a guide to the eye. The junction transparency is $\mathcal{D}_0=0.1$. In
the inset the zero-voltage value of the differential shot noise is plotted as a
function of $q$ (c.f. the text), in the Born limit [squares - same data as in the
(b)] and unitary limit [circles - same data as in the (b) part of
Fig.~\ref{Unitary}].} \label{Born}
\end{figure}

In Fig.~\ref{Born} and Fig.~\ref{Unitary} we plot the differential shot
noise for several pair breaking parameters, $\Gamma$, for scattering in
both the Born ($\sigma\ll 1$) and the unitary limits ($\sigma=1$),
respectively. Impurity renormalization leads to broadening of
quasiparticle states that depends on the pair breaking parameter,
$\Gamma$, and the scattering cross section, $\sigma$.

The local self energy at the interface is different from that in the bulk because of
the formation of  surface bound states. In particular, the surface bound state has a
large impurity renormalization in the Born limit, but is weakly modified in the
unitary limit.\cite{poe99} This is opposite to the state of affairs in the bulk,
where scattering in the unitary limit is more detrimental to the $d_{x^2-y^2}$ order
parameter, and produces a
low-energy impurity band in the density of states.

\begin{figure}[t]
\includegraphics[width=8cm]{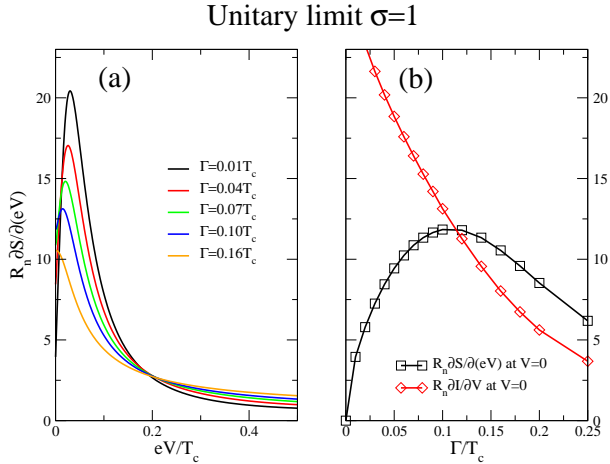}
\caption{\small The same as in Fig.~\ref{Born} but for scattering in the unitary limit
($\sigma=1$).}\label{Unitary}
\end{figure}

Impurity broadening of the surface ABS reduces the resonant transmission in the Andreev
channel, but opens up the single-particle tunnelling channel. In Fig.~\ref{Unitary_scatt}
we plot the reflection and transmission probabilities for scattering in the unitary limit
for several values of the pair breaking parameter. With increasing $\Gamma$, the reduction
of the Andreev reflection probability $R^R_{he}$ (Fig.~\ref{Unitary_scatt}a) is
accompanied by an increase of the normal reflection probability $R^R_{ee}$
(Fig.~\ref{Unitary_scatt}b) and an increase of the transmission probabilities, both
transmission without branch conversion $T^R_{ee}(1-|\tilde\gamma^R|^2)$
(Fig.~\ref{Unitary_scatt}c) and transmission with branch conversion
$T^R_{he}(1-|\gamma^R|^2)$ (Fig.~\ref{Unitary_scatt}d). In particular, the transmission
probabilities acquire a resonance form, similar to that in the Andreev reflection channel.

We note that probability is always conserved during scattering at the interface; it can be checked that
\begin{equation}\label{ProbCons}
R^R_{ee}+R^R_{he} +T^R_{ee}(1-|\tilde\gamma^R|^2)+T^R_{he}(1-|\gamma^R|^2) = 1.
\end{equation}
The third and fourth terms, which describe single particle
tunnelling, are identically zero in the sub-gap region in the
absence of impurity scattering, but become increasingly important
as the impurity renormalization increases (see
Fig.~\ref{Unitary_scatt}). When the Andreev resonance is reduced
and single-particle tunnelling becomes important, the
differential shot noise at zero voltage becomes non-zero, as
shown in Figs.~\ref{Born}-\ref{Unitary}. This is in line with the
phenomenological discussion in Ref.~\onlinecite{lof02a}. Thus, we
find that the noise-\textsl{less} character of the zero-energy
surface bound state is quickly lost when intrinsic broadening is
present.

To quantitatively assess the importance of impurity scattering
in tunnelling, the contribution to the width of the bound state
from impurity broadening, which we denote $w_i$, has to be
compared with the contribution set by the transparency of the
interface, $w_b$, introduced in the previous section. The width
$w_i$ is related to the imaginary part of the impurity self
energy $\Sig^R_3$ near the surface. Unfortunately, a rigorous
analytic calculation of $w_i$ in which the spatial dependence
of the impurity self energy and the order parameter are taken
into account has so far not been carried out (see, however, the
scaling analysis in Ref.~\onlinecite{poe99}). We estimate the
width to be $w_i = c
|\mbox{Im}\{\Sig^R_3(\epsilon=0,z=0^+)\}|$, where $c$ is
numerical factor which corrects for the spatial dependence of
the self energies. In the limit $w_i \gg w_b$, which we call
the `tunnel limit', intrinsic broadening is large and only
single-particle tunnelling is important. Andreev reflection can
then be neglected and the shot noise for low transparency
reduces to the Schottky form $S=2eI$, and does not contain any
new information that can not be extracted from the current. On
the other hand, in the limit $ w_i \ll w_b $, which we call the
`Andreev limit', impurity broadening is negligible, single
particle tunnelling is suppressed and Andreev reflection is
resonant. In this limit the shot noise is non-trivial. In
Figs.~\ref{Born}-\ref{Unitary}, the crossover between these two
regimes is displayed for impurity scattering in the Born and
unitary limits. It is clear that the impurity renormalization
near the surface is much larger in the Born limit compared to
the unitary limit: the crossover appears for $\Gamma/T_c\sim
10^{-3}$ in the Born limit, which is two orders of magnitudes
smaller than in the unitary limit. However, if we plot
$R_n\partial S/\partial (eV)$ at $V=0$ as a function of $ q =
w_i / \langle w_b({\bf p}_f) \rangle_{{\bf p}_f} $, with the
numerically computed $\mbox{Im}\{\Sig^R_3(\epsilon=0,z=0^+)$
and an estimate $c=1/3$ in the Born limit and $c=3$ in the
unitary limit, we find that the crossover appears near $q\sim
1$ in both limits, see inset of Fig.~\ref{Born}(b).

\begin{figure}
\includegraphics[width=8cm]{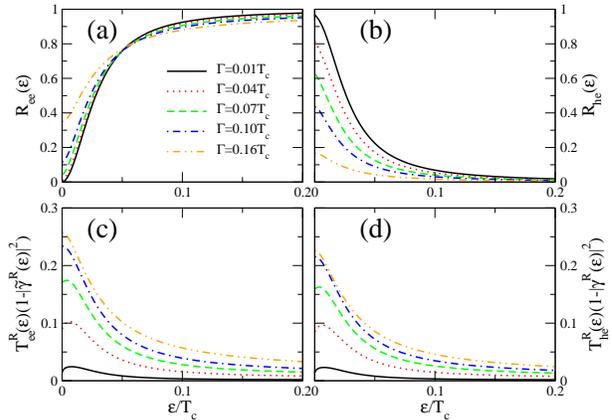}
\caption{\small Scattering probabilities for different pair
breaking strengths for tunnelling ranging from the the Andreev
limit (black curves) to the tunnel limit (brown curves). The
parameters corresponds to unitary scattering in
Fig.~\ref{Unitary}(a), and all probabilities were computed at an
incidence angle of $45^o$ relative to the interface normal. The
resonance in the Andreev reflection probability
$R_{he}(\epsilon)$ at zero energy (due to the bound state) is
broadened by impurity scattered and suppressed in the tunnel
limit. The resonance width in the Andreev limit is set by the
transparency of the interface. Note that the sum of all
probabilities is always equal to one, c.f. Eq.~(\ref{ProbCons}).}
\label{Unitary_scatt}
\end{figure}

\subsection{$d_{x^2-y^2}+is$ and $d_{x^2-y^2}+id_{xy}$ symmetries}

Finally, we consider the signatures of a surface phase transition from an
inhomogeneous $d_{x^2-y^2}$ surface phase, to a surface state with mixed symmetry:
$d_{x^2-y^2}+is$ or $d_{x^2-y^2}+id_{xy}$. In the clean limit the noise spectrum is
sensitive to the change of the surface excitation spectrum induced by the sub-dominant
pairing channel. When a complex order parameter develops near the surface,
time-reversed partners of the two-fold degenerate zero-energy bound states are
shifted in opposite directions from the Fermi level. The positive energy bound
state spectra for these mixed-symmetry phases are shown in Fig. (\ref{Surface_OP}).
A surface current, and an associated spontaneous magnetic field, are generated.
This symmetry breaking can be detected in the conductance as a spontaneous
splitting of the zero-bias conductance peak, or as a spontaneous
magnetic signal from the surface.

In the shot noise, for a clean system, we thus expect the point of vanishing noise to
disperse with angle of incidence, in a similar way the Doppler shift changes the shot
noise spectrum in an applied field. In addition to the dispersion of the bound states
with angle, there is an additional mechanism of dispersion due to electron-hole
de-phasing, which appears at non-zero energies when the order parameter has a spatial
dependence (pairbreaking suppression near the surface). These mechanisms of
dispersion of the bound state conspire to wash out much of the structure one might
otherwise expect to observe in the shot noise near zero voltage. Nevertheless, for a
clean superconductor, characteristic differences can be seen in $\partial
S/\partial V$
for the cases of sub-dominant $s$- and $d_{xy}$ order parameters.

\begin{figure}[t]
\includegraphics[width=8cm]{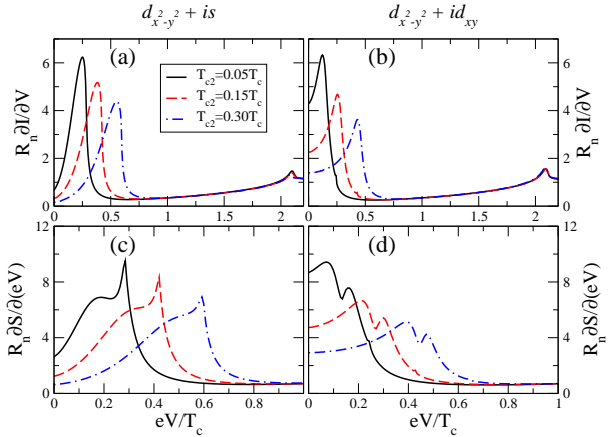}
\caption{\small (a)-(b) Conductance and (c)-(d) differential shot noise for an order parameter with an $s$
subdominant component (left column) and a $d_{xy}$ subdominant component (right column) for several different
interaction strengths. The barrier transparency is $\mathcal{D}_0=0.1$, and the system is in the Andreev limit
($\Gamma=0.001T_c$, $\sigma=10^{-4}$).}\label{subs}
\end{figure}

In Fig.~\ref{subs} we plot the conductance and differential shot noise for several different interaction
strengths, in both the $s$ and $d_{xy}$ sub-dominant channels. There is not difference in the signatures of
the two different subdominant components in the conductance: in both the $s$ and $d_{xy}$ cases the zero-bias
conductance peak is split and appears at a finite voltage related to the size of the subdominant order
parameter. On the other hand, in the shot noise, there is a double peak structure in both cases, with the
high-voltage peak bigger than the low-voltage peak for the $s$-wave case, but with a reversal in spectral
weight between low- and high-voltage peaks for the $d_{xy}$ case. This reversal reflects the difference in the
dispersion of the bound states for the two different pairing channels, which affects the point of
suppressed shot noise (as well as the associated peak in $\partial S/\partial V$). The shift to finite voltage
is larger and disperses less in the $s$-wave case, compared to the $d_{xy}$ case.

Application of a magnetic field introduces additional dispersion, and the angle integration leads to a
reduction of the structures. For both types of order parameters, the double peak evolves with increasing
magnetic field strength into a single peak on a scale $H/H_0$ set by the size of the subdominant gap, see
Fig.~\ref{subs_dep}(a).

As in the pure $d$-wave case discussed in the previous section, impurity scattering broaden the bound state
resonance and reduces the structure in the shot noise. Thus, with increasing pair breaking parameter the two
peaks in the shot noise merge into a single peak and, in the tunnel limit, the differential shot noise reduces
to the conductance, see Fig.~\ref{subs_dep}(b).

\begin{figure}[t]
\includegraphics[width=8cm]{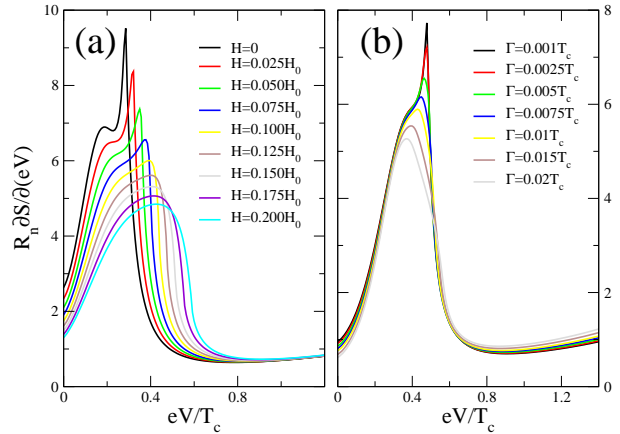}
\caption{(a) Magnetic field dependence of differential shot noise for
 a $d_{x^2-y^2}+is$ order parameter with $T_{c2}=0.05T_c$. Here is
 $\Gamma=0.001T_c$ and $\sigma=10^{-4}$. (b) Differential shot noise
 as a function of pair breaking parameter $\Gamma$ in the Born limit
 ($\sigma=10^{-4}$). Here is $T_{c2}=0.2T_c$. In both cases is
 $\mathcal D_0=0.1$.}\label{subs_dep}
\end{figure}

\section{Discussion and Summary}\label{SecSum}

In conclusion, shot noise can be a useful tool to extract detailed information about properties of junctions
between normal metals and unconventional superconductors. However, a necessary condition is that the system is
in the Andreev limit, $w_i\ll w_b$, as shown in Figs.~\ref{Born}-\ref{Unitary_scatt}. This is a rather
restrictive condition at present, since in most experiments the zero-bias conductance peak is broadened by
disorder, see however Ref.~\onlinecite{wei98}.

In several recent experiments\cite{hud99,yaz99,pan00} the density of states around single impurities or
inhomogeneities were mapped out by $c$-axis STM spectroscopy. The results are discussed in terms of low energy
states bound to a single impurity scattering in the unitary limit, in line with theoretical works in
Refs.~\onlinecite{lee93,sal96,fla98}. Thus, if impurities in the high-$T_c$ superconductors are indeed
scattering in the unitary limit, we expect that the surface bound states will not be particularly broadened,
and tunnelling in the Andreev limit should be possible to achieve experimentally in a clean sample. The mean
free path corresponding to $q\ll 1$ in the unitary limit (for the parameters in Fig.~\ref{Unitary}) is
estimated to be of the order of tenths of coherence lengths, which is achievable experimentally.

There are other sources of broadening of the surface/interface bound states that were not considered in this
paper. In particular, it is clear that surface roughness will drive the system towards the tunnel limit,
because non-specular scattering of quasiparticles to the nodes of the order parameter broadens the bound
states just as impurity scattering does. Therefore, to extract information from shot noise it will be
important to have a specularly reflecting junction, or tunnel from an STM tip directly into the $ab$-plane of
a specular portion of a superconductor surface.

Under these circumstances, information about the superconductor properties can be deduced via the particular
properties of the zero energy surface bound states. The shot noise in a purely $d_{x^2-y^2}$-wave
superconductor is suppressed around low voltage and approaches zero in the clean limit. The characteristic
magnetic field dependence shown in Fig.~\ref{dSdV_Hext} can then be used to test the theory. The zero-voltage
shot noise level changes according to the inset of Fig.~\ref{dSdV_Hext}, and the satellite peak is linearly
pushed out with increasing magnetic field strength. The double peak structure shown in Fig.~\ref{subs} serves
as a fingerprint of the symmetry of the subdominant pairing channel.

\acknowledgments

We thank A. Vorontsov and M. Eschrig for valuable discussions. This work was supported by the NSF, grant DMR
9972087, the Swedish Research Council, VR, and STINT, The Swedish Foundation for International Cooperation in
Research and Higher Education.

\begin{center}
\begin{table*}
\caption{The Andreev reflection probabilities. The denominators are listed in Table~\ref{TRprobabilities}.}
\begin{tabular}{|l|}
\hline $r_{he} = \tilde \Gamma^R_1 = [R(1+\tilde\gamma^R_2\gamma^R_2)\tilde\gamma^R_1
+D(1+\tilde\gamma^R_1\gamma^R_2)\tilde\gamma^R_2]/\zeta_1$ \\
$r_{eh} = \Gamma^R_1 = [R(1+\gamma^R_2\tilde\gamma^R_2)\gamma^R_1
+D(1+\gamma^R_1\tilde\gamma^R_2)\gamma^R_2]/\zeta_2$ \\
$\bar r_{he} = \tilde\Gamma^R_2 = [R(1+\gamma^R_1\tilde\gamma^R_1)\tilde\gamma^R_2
+D(1+\tilde\gamma^R_2\gamma^R_1)\tilde\gamma^R_1]/\zeta_3$ \\
$\bar r_{eh} = \Gamma^R_2 = [R(1+\gamma^R_1\tilde\gamma^R_1)\gamma^R_2
+D(1+\gamma^R_2\tilde\gamma^R_1)\gamma^R_1]/\zeta_4$\\
\hline
\end{tabular}\label{ARamplitudes}
\end{table*}
\end{center}

\begin{center}
\begin{table*}
\caption{Scattering probabilities for $x^K$ distribution functions in the stationary SIS junction setup.}
\begin{tabular}{|l|l|l|l|}
\hline $\bar T_{hh} = D|1+\tilde\gamma^R_1\gamma^R_2|^2/|\zeta_1|^2$ & $\bar T_{he} =
RD|\tilde\gamma^R_1-\tilde\gamma^R_2|^2/|\zeta_1|^2$ & $R_{hh} =
R|1+\tilde\gamma^R_2\gamma^R_2|^2/|\zeta_1|^2$ &
$\zeta_1 = 1+R\gamma^R_2\tilde\gamma^R_2+D\tilde\gamma^R_1\gamma^R_2$\\
\hline $\bar T_{ee} = D|1+\gamma^R_1\tilde\gamma^R_2|^2/|\zeta_2|^2$ & $\bar T_{eh} =
RD|\gamma^R_1-\gamma^R_2|^2/|\zeta_2|^2$ & $R_{ee} = R|1+\gamma^R_2\tilde\gamma^R_2|^2/|\zeta_2|^2$ &
$\zeta_2 = 1+R\gamma^R_2\tilde\gamma^R_2+D\gamma^R_1\tilde\gamma^R_2$ \\
\hline $T_{hh} = D|1+\tilde\gamma^R_2\gamma^R_1|^2/|\zeta_3|^2$ & $T_{he} =
RD|\tilde\gamma^R_2-\tilde\gamma^R_1|^2/|\zeta_3|^2$ & $\bar R_{hh} =
R|1+\gamma^R_1\tilde\gamma^R_1|^2/|\zeta_3|^2$ &
$\zeta_3 = 1+R\gamma^R_1\tilde\gamma^R_1+D\tilde\gamma^R_2\gamma^R_1$ \\
\hline $T_{ee} = D|1+\gamma^R_2\tilde\gamma^R_1|^2/|\zeta_4|^2$ & $T_{eh} =
RD|\gamma^R_2-\gamma^R_1|^2/|\zeta_4|^2$ & $\bar R_{ee} = R|1+\gamma^R_1\tilde\gamma^R_1|^2/|\zeta_4|^2$ &
$\zeta_4 = 1+R\gamma^R_1\tilde\gamma^R_1+D\gamma^R_2\tilde\gamma^R_1$ \\
\hline
\end{tabular}\label{TRprobabilities}
\end{table*}
\end{center}

\section*{Appendix}

In this appendix we tabulate the Andreev reflection probabilities expressed in terms of the barrier reflection
and transmission probabilities and coherence amplitudes for incoming trajectories in Table \ref{ARamplitudes},
and in Table \ref{TRprobabilities} we summarize the transmission and reflection probabilities that enter the
boundary conditions for the distribution functions in Eqs. (\ref{x_bc1}-\ref{x_bc4}).


\end{document}